\documentclass[pra,
superscriptaddress,
amssymb,amsmath,amsmath,showpacs,reprint,twocolumn]{revtex4-1}
\usepackage{color}
\usepackage{graphicx}
\usepackage{epstopdf}
\usepackage{natbib}
\usepackage{wasysym}
\usepackage{hyperref}
\usepackage{dcolumn}

\hypersetup{%
   pdfpagemode=None, 
   pdfstartpage=1,
   pdfstartview=FitH,
   pdfmenubar=true,
   pdftoolbar=true,
   colorlinks = true,
   linkcolor=blue,
   citecolor=blue,
   bookmarksopen=false
 }

\newcommand{\Fkt}[1]{\,\mathsf {#1}}

\def\openone{\leavevmode\hbox{\small1\kern-3.3pt\normalsize1}}

\ifx\Tr\renewcommand{\Tr}{\Fkt{Tr}} 
\else\newcommand{\Tr}{\Fkt{Tr}}
\fi

\usepackage{booktabs}
\usepackage{graphicx}
\usepackage{amsmath}
\usepackage{indentfirst}
\begin{document}
\title{Calculation of two-centre two-electron integrals over Slater-type orbitals revisited. \\III.
Case study of the beryllium dimer}

\author{\sc Micha\l\ Lesiuk}
\email{e-mail: lesiuk@tiger.chem.uw.edu.pl}
\author{\sc Micha\l\ Przybytek}
\affiliation{\sl Faculty of Chemistry, University of Warsaw, Pasteura 1, 02-093 Warsaw, Poland}
\author{\sc Monika Musia\l}
\affiliation{\sl Institute of Chemistry, University of Silesia, Szkolna 9, 40-006 Katowice, Poland}
\author{\sc Bogumi\l\ Jeziorski}
\author{\sc Robert Moszynski}
\affiliation{\sl Faculty of Chemistry, University of Warsaw, Pasteura 1, 02-093 Warsaw, Poland}
\date{\today}

\begin{abstract}
In this paper we present results of \emph{ab-initio} calculations for the beryllium dimer with basis set of Slater-type 
orbitals (STOs). Nonrelativistic interaction energy of the system is 
determined using the frozen-core full configuration interaction calculations combined with high-level coupled cluster 
correction for inner-shell effects. Newly developed STOs basis sets, ranging in quality from double to sextuple 
zeta, are used in these computations. Principles of their construction are discussed and several atomic benchmarks are 
presented. Relativistic effects of order $\alpha^2$ are calculated perturbatively by using the Breit-Pauli 
Hamiltonian and are found to be significant. We also estimate the leading-order QED 
effects. Influence of the adiabatic correction is found to be negligible. Finally, the interaction energy of the
beryllium dimer is determined to be 929.0$\,\pm\,$1.9 
$\mbox{cm}^{-1}$, in a very good agreement with the recent experimental value. The results presented here 
appear to be the most accurate \emph{ab-initio} calculations for the beryllium dimer available in the literature up to 
date and probably also one of the most accurate calculations for molecular systems containing more than four electrons.
\end{abstract}

\maketitle

\section{Introduction}
\label{sec:intro}
State-of-the-art \emph{ab initio} electronic structure calculations are very important for the new emerging field at the
border of chemistry and physics - the studies of ultracold molecules. During the past decades, experimental advances in
laser cooling and trapping of neutral atoms have opened a door for the formation of ultracold diatomic molecules by
photoassociation \cite{julienne06a} and magnetoassociation \cite{julienne06b} techniques. In this respect, \emph{ab
initio} calculations of the potential energy curves and coupling matrix elements between the electronic states turned
out to be crucial to interpret the experimental observations. See, for instance, Ref. \cite{mcguyer13} for the
theoretical explanation of the unusual quadratic Zeeman shifts in the Sr$_2$ molecule, or Ref. \cite{mcguyer14} for
interpretation of the observed subradiant states of Sr$_2$. Electronic structure calculations can also be used to
predict new schemes for the formation of ultracold diatomic molecules
\cite{tomza13a,krych11,skomorowski12,tomza13b,tomza14}. Apart from that, state-of-the-art first-principles calculations
are used in metrology \emph{e.g.} to determine the pressure standard \cite{jeziorski10}. Last but not least, accurate
interatomic interaction potentials are of significant importance in search for a new physics. See \emph{e.g.} Ref.
\cite{jeziorski12} for a theoretical study of the QED retardation effect of the helium dimer, and the work of Zelevinsky
\emph{et al.} \cite{zelevinsky08} for a joint experimental-theoretical efforts towards determination of the
proton-electron mass ratio time variation. Additionally, one can mention the work of Schwertweger \emph{et al.}
\cite{schwertweger11} on the Sr$_2$ molecule where time variation of the fine structure constant is investigated.

All the aforementioned physical applications require high-precision theoretical data. Slater-type orbitals (STOs) are
expected to improve the description of many-electron systems, thus leading to results more accurate than available
at present. In the first two papers of the series we have proposed new efficient algorithms for the calculation of
two-centre integrals over STOs. As the first application of the STOs integral code we performed calculations for the
beryllium dimer in its ground $^1\Sigma_g^+$ state. This is a
challenging system, both from the theoretical and experimental point of view. From the theory side, it has already been
known that in order to reach accurate results very advanced quantum chemistry methods must be used. In fact, probably
the first calculations performed for this system 
by Fraga and Ransil \cite{fraga61}, using the resticted Hartree-Fock (RHF) method, led to the conclusion that the 
potential energy curve is purely repulsive. Further inclusion of the electron correlation, by using the configuration 
interaction method (CI) with single and double substitutions (CISD), appeared to confirm this 
observation \cite{bender67}. However, more refined calculations with the same method indicated an existence of a weak 
bond \cite{dykstra76,blomberg78}, with the interaction energy of order of several tens of $\mbox{cm}^{-1}$ and 
equilibrium distance of $\approx 5$ \AA{}, which is characteristic for the van der Waals molecules such as 
$\mbox{Ne}_2$. A similar conclusion was found in a study \cite{chiles81} employing the coupled cluster (CC) 
methods with double (and single) excitations (CCD, CCSD).

However, somehow later Harrison and Handy \cite{harrison83} performed frozen-core full configuration interaction (FCI) 
calculations 
and found that the interaction energy is at least several hundreds of $\mbox{cm}^{-1}$ larger. Even more importantly, 
they reported the presence of a deep minimum around $2.5$ \AA{} which was a rather unexpected result at this time. 
These results indicate that the connected triple (and possibly also quadruple) excitations are 
responsible for the formation of the bond. Reasons for such slow convergence of the traditional configuration 
interaction or coupled cluster expansions were analysed in details by Liu \emph{et al.} \cite{liu80}. It was shown that 
the pathological behaviour of this system encountered during studies performed with the 
single reference methods is mostly due to near-degeneracy of the $2s$ and $2p$ orbitals of the beryllium atom. It gives 
the beryllium dimer a strongly multireference nature. By applying the multireference configuration interaction (MRCI) 
method, Liu \emph{et al.} found the interaction energy to be as large as 810 $\mbox{cm}^{-1}$ and confirmed the 
existence of the minimum around $2.5$ \AA{}. These findings were later verified by several independent MRCI
studies \cite{fusti96a,fusti96b,starck96,keledin99,schmidt10,mitin11,bausch92,khatib14}. Therefore, it is now well
established that Be$_2$ is \emph{not} a van der Waals molecule.

Since now, a large number of theoretical works entirely devoted to study of the beryllium dimer have been published and 
a more detailed bibliography is given elsewhere \cite{roeggen05,patkowski07}. The interaction energy is typically
determined to be within the range of 200-1000 $\mbox{cm}^{-1}$ and it varies with the level of theory and quality of the
basis sets used. However, it appears that in the most recent, and probably the most accurate, studies, the interaction
energy fluctuates somewhere around 900 $\mbox{cm}^{-1}$. For instance, Martin \cite{martin99} found 944$\,\pm\,$25
$\mbox{cm}^{-1}$, Gdanitz \cite{gdanitz99} - 989$\,\pm\,$8 $\mbox{cm}^{-1}$, Pecul \emph{et al.} \cite{pecul00} - 885
$\mbox{cm}^{-1}$, R\o{}ggen and Veseth \cite{roeggen05} - 945$\,\pm\,$15 $\mbox{cm}^{-1}$, Patkowski \emph{et al.}
\cite{patkowski07} - 938$\,\pm\,$15 $\mbox{cm}^{-1}$, Koput \cite{koput11} - 935$\,\pm\,$10 $\mbox{cm}^{-1}$, and 
Sharma \emph{et al.} - 931.2 $\mbox{cm}^{-1}$ \cite{sharma14}. Discrepancies between these results are still rather 
large, though, which indicates that the ground state of the beryllium dimer remains to be a challenge for modern 
quantum chemistry methods.

From the experimental point of view, the ground state of the beryllium dimer is also a demanding system. First 
empirical confirmation of the fact that Be$_2$ is a deeply bound system, as theoretically predicted, was reported in 
the eighties \cite{bondybey84a,bondybey84b,bondybey85}. The most frequently cited experimental result for the
well-depth was given by Bondybey \emph{et al.}, 790$\,\pm\,$30 $\mbox{cm}^{-1}$. This result is not accurate and
the true error is much larger than the estimated error bars. However, the discrepancy is not really due to the
experimental error but mostly due to theoretical assumptions used to extract the dissociation energy. In fact, in 2006
Spirko \cite{spirko06} combined the experimental data of Bondybey with the best theoretical potential energy curve
available at the time and refined the result to 923 $\mbox{cm}^{-1}$ which is much closer to the recent theoretical
findings. In 2009 a new experiment was performed by Merritt \emph{et al.} \cite{merritt09} and the interaction energy
was found to be 929.7$\,\pm\,$2.0 $\mbox{cm}^{-1}$. Additionally, eleven vibrational levels were characterised
\cite{bernath09}. Shortly afterwards, Patkowski \emph{et al.} \cite{patkowski09} suggested the existence of the twelfth
vibrational level, just 0.44 $\mbox{cm}^{-1}$ below the dissociation limit, by using the ``morphed'' theoretical
potential energy curves. 

It is clear that the ground state of the beryllium dimer is a challenging system, with large 
requirements for the quality of the basis set and for the theoretical methods. Therefore, it is a good test case for 
the Slater-type orbitals (STOs) combined with the state-of-the-art quantum chemistry methods. It is well-known that
STOs are able to satisfy the electron-nucleus cusp condition, thereby significantly improving the description of the
wavefunction in a vicinity of the nuclei. This property makes STOs more reliable in calculations which depend crucially
on the quality of the trial wavefunction in this regime, such as core-core and core-valence correlation effects,
one-electron relativistic corrections of order $\alpha^2$ \emph{etc.} Other advantages of STOs are summarised at the
end of the present paper. Notably, calculations with STOs basis sets of quality up to sextuple zeta, aiming at
spectroscopic accuracy, have never been 
performed thus far. In the case of such calculations special attention must be paid to technical issues, 
such as creation and benchmarking of basis sets, since the strategies adopted in case of Gaussian-type orbitals (GTOs) 
may not be straightforwardly transferable. In this communication we consider these issues in some detail but restrict
ourselves to calculations at the equilibrium internuclear distance, $R$, equal to $2.4536$ \AA{} which is the recent
experimental value \cite{merritt09}. The whole potential energy curve will be reported later, along with a detailed
study of the related spectroscopical issues.

This paper is organised as follows. In Section \ref{sec:pre} we describe in details the systematic construction of the 
STOs basis sets. In Section \ref{sec:bench} we present 
benchmarks for the beryllium atom which verify the reliability of the developed STOs basis sets. Issues connected 
with extrapolations towards the complete basis set (CBS) are also investigated. In Section \ref{sec:dimer} we present 
results for the ground state of the beryllium dimer. We calculate the valence and core correlations effects separately 
and estimate the corresponding errors. Additionally, we compute the values of the relativistic corrections and estimate 
the effects of the leading-order QED contributions. Finally, in Section \ref{sec:conclusion} we conclude the paper and 
give a short outlook.

\section{Basis sets}
\label{sec:pre}
In the case of Gaussian-type orbitals (GTO), the contracted functions are typically used to reproduce the Hartee-Fock 
energy first.
Then, additional uncontracted functions are used to describe the electronic correlation, see the works of Dunning
\cite{dunning89,dunning92,dunning93,dunning94,dunning95,dunning96,dunning99,dunning01,dunning11} as a
representative example. We found that GTO basis sets designed according to this principle somewhat lack flexibility for
the $l=0$ partial wave, especially in the molecular
environment,
since the number of uncontracted $1s$ orbitals is typically small. For ordinary GTO calculations this is not a problem,
however, because correlation energy retrieved by $l=0$ angular momentum functions is small - at least an order of
magnitude below the contribution from $l=1$ partial wave. Therefore, this lack of correlation coming from $l=0$
functions is visible only for very accurate calculations where the contributions from more important partial waves are
already sufficiently saturated. Since we aim at high quality results, we do not use contractions of STOs.

There is also another important choice in the design of STO basis sets which is entirely absent in the case of GTO. For 
GTO calculations one typically uses only $1s$, $2p$, $3d$ \emph{etc.} functions (with $n=l+1$) since molecular 
integrals with 
these kind of functions are particularly straightforward. In the case of STO one can use functions with $n>l+1$ as 
well. For instance, in the case of $l=0$ orbitals the expansion takes the following form
\begin{align}
\label{hyl}
 \psi_i = e^{-\zeta_i r} \sum_k^{N_i} c_{ki} \,r^k,
\end{align}
where the value of $\zeta_i$ is characteristic for a given atomic shell.
The expansion (\ref{hyl}) is quite attractive, mainly because of a small number of nonlinear parameters 
which need to be optimised - only one per atomic shell, and very systematic enlargement towards the completeness 
through the parameters $N_i$. However, in practice we found that there are numerous problems connected with 
this expansion in our applications. The biggest drawback is the fact that basis sets constructed according 
to the principle (\ref{hyl}) suffer from near-linear dependencies when $N_i$ gets moderate or large. This effectively 
prohibits the construction of large basis sets close to completeness when the standard double precision arithmetic is 
used. Another problem is the fact that the expansion (\ref{hyl}) is not as flexible 
as necessary, especially when transferred from atomic to a weakly bound molecular system.

As a result, we found that more flexible and well-behaved basis sets can be obtained when the orbitals are expanded, 
similarly as for GTO, in a set of functions with $n=l+1$ and their respective exponents are varied freely \emph{i.e.}
\begin{align}
\label{float}
 \psi_i = \sum_k^{N_i} c_{ki} \,e^{-\zeta_{ik} r}.
\end{align}
This choice, however, brings up the problem of optimisation of a large number of independent parameters $\zeta_{ik}$. 
In the biggest basis set created in this work a direct use of Eq. (\ref{float}) would require free optimisation of 
several tens of the nonlinear parameters. This is possible but very time consuming. Even more daunting problem
is the presence of a great number of local minima. There is no guarantee that a brute-force optimisation would have 
found the true global minimum, even with a decent starting point. This fact puts the reliability of the extrapolation 
towards the complete basis set (CBS) in question.

Aware of all the aforementioned issues, we adopted the strategy of \emph{even-termpering} so that the nonlinear 
parameters for a given angular momentum $l$ are in the following form
\begin{align}
\label{et}
\zeta_{lk} = \alpha_l\, \beta_l^k \;\;\; \mbox{with}\; k=0,1,2,\ldots
\end{align}
Nowadays, even-termpering is routinely applied for construction of GTOs basis sets. However, this technique was 
originally proposed by Raffenetti and co-workers \cite{raffenetti72,raffenetti73} in the context of STOs. 
Even-termpering greatly reduces the number of independent parameters which need to be optimised 
(only two for each partial wave).

\begin{table*}[ht]
\caption{Composition of STO basis sets for the beryllium atom.}
\begin{ruledtabular}
\begin{tabular}{c|lll}
\label{base}
basis set & atomic valence & tight core & diffuse \\
\hline\\[-2.1ex]
ATC-ETCC-1 & $6s1p$ & $1s$ & $1s1p$ \\[0.6ex]
ATC-ETCC-2 & $7s2p1d$ & $1s1p$ & $1s1p1d$ \\[0.6ex]
ATC-ETCC-3 & $8s3p2d1f$ & $2s2p1d$ & $1s1p1d1f$ \\[0.6ex]
ATC-ETCC-4 & $9s4p3d2f1g$ & $2s3p2d1f$ & $1s1p1d1f1g$ \\[0.6ex]
ATC-ETCC-5 & $9s5p4d3f2g1h$ & $3s4p3d2f1g$ & $1s1p1d1f1g1h$ \\[0.6ex]
ATC-ETCC-6 & $9s6p5d4f3g2h1i$ & $3s5p4d3f2g1h$ & $1s1p1d1f1g1h1i$ \\[0.6ex]
\end{tabular}
\end{ruledtabular}
\end{table*}

The first step in the creation of the STOs basis sets is optimisation of the atomic valence
basis set. In this step the core $1s$ orbital of the beryllium atom is kept frozen and CISD method, 
equivalent to FCI
for the valence shell, is used. The optimisation is carried out to minimise the \emph{total} energy of the two-electron
CISD \emph{i.e.} sum of the Hartree-Fock and CISD correlation energy.

Since the seminal work of Dunning and co-workers 
\cite{dunning89,dunning92,dunning93,dunning94,dunning95,dunning96,dunning99,dunning01,dunning11} it has been known that 
to allow for a reliable extrapolation towards CBS, basis sets need to be constructed 
according to the \emph{correlation consistency} principle. Roughly speaking, it ensures that at a given stage all 
functions which give approximately the same energy contributions are simultaneously included. Our atomic valence basis 
sets are denoted ETCC-L which stands for even-tempered correlation consistent and L is the largest angular momentum 
included. Therefore, ETCC-1 has the composition $6s1p$, ETCC-2 - $7s2p1d$ and so forth, and only functions with $n=l+1$ 
are used. The initial number of six $1s$ functions was found to be optimal. Compositions of all basis sets up to
$L=6$ are 
presented in Table \ref{base}. At some point it becomes 
unnecessary to include more $1s$ functions, and thereafter their number was kept fixed. The even-tempered expansion, 
(\ref{et}), is used separately for each partial wave.

The second step in construction of the basis set for beryllium is addition of the ``tight'' functions which are 
necessary for description of the core-core and core-valence correlations. It is well-known that the core electrons are 
chemically inert and their contribution to the total energy cancels out to a large extent when interaction energies are 
computed. This observation is the foundation for the so-called frozen core approximation. However, in accurate 
calculations the frozen core approximation cannot be applied, especially for an element such as beryllium. Obviously, 
valence basis sets cannot describe the core-core and core-valence correlations since polarisation functions with large 
exponents, characteristic for the core, are absent. We added core polarisation functions to the previously obtained 
ETCC-L basis sets. Detailed composition of the extended TC-ETCC-L basis sets (where TC stands for ``tight core'') is 
given in Table \ref{base} for each L. In order to optimise the exponents of the core polarisation functions we 
minimised the difference between the total energies of all-electron CISD and frozen-core CISD for the beryllium atom. 
Since the number of independent nonlinear parameters was much smaller than for the valence basis sets, even-tempering 
of the exponents was not necessary and all variables were optimised freely. A minor detail of the optimisation 
procedure is that the derivative of the target function with respect to the logarithm of the exponent was used as a 
gradient, rather than the derivative with respect to the exponent itself. This stabilises greatly the 
numerical performance of the optimisation.

The third, and final, step of the basis sets creation is the addition of the diffuse functions. These functions are not 
necessary for the atomic calculations since tails of 
the electron density do not contribute greatly to the total energies of the atom. However, in a molecular 
environment tails of the electron density are responsible for the act of bonding in weakly interacting systems and 
accurate reproduction of the potential energy curve. Basis sets augmented with a set of diffuse functions are called 
A-ETCC-L, or ATC-ETCC-L in the case of the core-valence basis sets. Detailed structure of the augmented basis sets is 
given in Table \ref{base}. Exponents of the diffuse functions were optimised to maximise the absolute value of the
beryllium dimer interaction energy calculated with A-ETCC-L basis sets at four electron (valence) CCSD(T) level of
theory \cite{raghavachari89}.

Notably, the strategy that the diffuse functions are optimised to maximise the absolute value of the interaction energy
makes them formally dependent on the internuclear distance, $R$. This is, in fact, exactly in line with our intentions.
In this work we consider only one value of $R$, corresponding to the minimum of the potential energy curve, so that
there is no 
ambiguity in how the calculations are carried out. In case when a complete potential energy curve is required, diffuse 
functions can be optimised for several values of $R$ and then interpolated smoothly. 
The present approach is inspired by the works of Ko\l os and co-workers concerning the hydrogen molecule 
\cite{kolos60a,kolos60b,kolos64,kolos68}. Basis sets used in these works contained several nonlinear parameters which 
were handled similarly as described above and no significant difficulties were reported.

All optimisations necessary to construct the basis sets were carried out by using pseudo Newton-Rhapson method with 
the BFGS update of the approximate Hessian matrix \cite{fletcher81}. Our own code, written especially for this purpose, 
was used throughout. This program is interfaced with the \textsc{Gamess} package \cite{gamess1,gamess2} which carries 
out the electronic structure calculations. Gradient with respect to nonlinear parameters was calculated numerically 
with the two-point finite difference formula. Close to a minimum, where more accurate values of the gradient are 
necessary, the four-point finite difference formula was applied. Optimisation was stopped when the energy differences 
between two consecutive iterations fell below 1 nH and the largest element of the gradient below 10 $\mu$H, 
simultaneously. Typically, several tens of iterations were necessary to converge to a minimum in the biggest 
calculations. To avoid the exponent values of two functions to collapse, which occasionally happened, a Gaussian-type 
penalty function was applied routinely.

STOs constitute a convenient basis set for calculation of the relativistic corrections because of the cusp at the
origin. Nonetheless, it is obvious that standard STO basis sets used in calculation of the Born-Oppenheimer potential
may not be fully satisfactory. To overcome this problem we modified our ATC-ETCC-L basis sets by replacing all $1s$ 
orbitals by a new set, common for each L. The latter consists of fifteen functions and was trained to minimise the
Hartree-Fock energy
of the beryllium atom. The value obtained, $-14.5730231385$, differs at 10th significant digit from the best estimate
available in the literature, $-14.573023168305$ \cite{kobus12}. The S-extended basis sets are abbreviated shortly 
ATC-ETCC-L+S.

Composition of the STO basis sets along with detailed values of the exponents and quantum numbers are given in the
Supplementary Material \cite{supplement}.

\section{Atomic benchmarks}
\label{sec:bench}

\subsection{Nonrelativistic energy}
\label{subsec:boatom}
The beryllium atom is a convenient system for benchmarking purposes because accurate reference values of the total 
energies and relativistic corrections are available in the literature. Therefore, before the
calculations on the diatomic system are given, it is useful to check the adequacy of the strategy and the performance 
of our basis sets in the atomic case. We calculated the full configuration interaction (FCI) energies of the beryllium 
atom by using ATC-ETCC-L basis sets with $L$ = 2,...6. Newly developed, general FCI program \textsc{Hector} 
\cite{przybytek14}, written 
by 
one of us (M.P.), was used for this purpose. The starting Hartree-Fock orbitals were taken from the \textsc{Gamess} 
program package, interfaced with our STO integral code.

In Table \ref{coratom} we present the FCI results for the beryllium atom. It is important for further developments to
extrapolate these results towards the complete basis set. Many extrapolation methods were suggested in the literature 
\cite{feller92,peterson94,martin96,halkier98}, but the following formula was found to be particularly reliable for 
estimation of the CBS limit of the correlation energy
\begin{align}
\label{extra}
E = A + \frac{B}{L^3} + \frac{C}{L^5},
\end{align}
where $L$ is the largest angular momentum present in the basis set. The Hartree-Fock results were not extrapolated but
simply the value in the biggest basis set was taken. 
Extrapolation of the results given in Table \ref{coratom} leads to the result $-14.667 345$ for the total energy of the
beryllium atom. This can be compared with the reference value, obtained by Pachucki and Komasa \cite{pachucki04a} by 
using explicitly correlated four-electron basis set, 
$-14.667 356$, and the error is equal to $11\mu$H. Remarkably, the extrapolation reduces the error by an order of 
magnitude, compared with the largest basis set available. In fact, we found that an essential feature of STOs basis 
sets is that they provide very reliable extrapolation towards the CBS limit, as compared with GTOs basis sets of a 
similar quality. 

\begin{table}[ht]
\caption{Total energy, $E_{total}$, and the correlation energy, $E_c$, of the beryllium atom calculated at the FCI level
of theory by using the STOs basis sets ATC-ETCC-L. The limit of the Hartree-Fock energy is assumed to be $-$14.573023 
H.}
\begin{ruledtabular}
\label{coratom}
\begin{tabular}{cccc}
basis set & $E_c$/mH & $E_{total}$/H \\
\hline\\[-2.1ex]
ATC-ETCC-2 & $-$85.976 & $-$14.658 998 \\[0.6ex]
ATC-ETCC-3 & $-$91.479 & $-$14.664 502 \\[0.6ex]
ATC-ETCC-4 & $-$92.994 & $-$14.666 017 \\[0.6ex]
ATC-ETCC-5 & $-$93.608 & $-$14.666 631 \\[0.6ex]
ATC-ETCC-6 & $-$93.902 & $-$14.666 925 \\[0.6ex]
\hline\\[-2.1ex]
CBS & $-$94.322 & $-$14.667 345 \\[0.6ex]
\hline\\[-2.1ex]
Pachucki and Komasa \cite{pachucki04a}  & $-$94.333 & $-$14.667 356 \\
\end{tabular}
\end{ruledtabular}
\end{table}

\subsection{One-electron relativistic corrections}
\label{subsec:relatom}

\begin{table}[ht]
\caption{Mass-velocity, $\langle P_4 \rangle$, and one-electron Darwin, $\langle D_1 \rangle$, corrections for the
beryllium atom at the FCI level of theory. The factor of $\alpha^2$ is not included. All values are given in the atomic
units.}
\begin{ruledtabular}
\label{relatom}
\begin{tabular}{cccc}
basis set & $\langle P_4 \rangle$ & $\langle D_1 \rangle$ \\
\hline\\[-2.1ex]
ATC-ETCC-2+S & $-$270.431 854 & 222.218 606 \\[0.6ex]
ATC-ETCC-3+S & $-$270.527 702 & 222.225 660 \\[0.6ex]
ATC-ETCC-4+S & $-$270.568 886 & 222.232 142 \\[0.6ex]
ATC-ETCC-5+S & $-$270.594 238 & 222.234 514 \\[0.6ex]
ATC-ETCC-6+S & $-$270.609 955 & 222.235 299 \\[0.6ex]
\hline\\[-2.1ex]
CBS & $-$270.648 568 & 222.236 568 \\[0.6ex]
\hline\\[-2.1ex]
Pachucki and Komasa \cite{pachucki04a}  & $-$270.704 68(25) & 222.229 35(13) \\
\end{tabular}
\end{ruledtabular}
\end{table}

The leading relativistic corrections (the second order in the fine structure constant, $\alpha$) to the energy 
of light systems can be computed perturbatively as an expectation value of the Breit-Pauli Hamiltonian \cite{bethe75}.
For a molecule in a singlet state this correction is \cite{pachucki04b,pachucki07}
\begin{align}
\label{breit}
E^{(2)} &= \langle P_4 \rangle + \langle D_1 \rangle + \langle D_2 \rangle + \langle B \rangle,\\
\langle P_4 \rangle &= -\frac{\alpha^2}{8} \langle \sum_i \nabla_i^4 \rangle,\\
\langle D_1 \rangle &= \frac{\pi}{2}\alpha^2\sum_a Z_a \langle \sum_i \delta(\textbf{r}_{ia})\rangle,\\
\langle D_2 \rangle &= \pi\alpha^2\langle\sum_{i>j}\delta(\textbf{r}_{ij})\rangle,\\
\langle B \rangle   &= \frac{\alpha^2}{2}\langle \sum_{i>j} \left[\frac{\nabla_i\cdot\nabla_j}{r_{ij}}
+\frac{\textbf{r}_{ij}\cdot(\textbf{r}_{ij}\cdot\nabla_j)\nabla_i}{r_{ij}^3}\right] \rangle,
\end{align}
where $\langle \hat{\mathcal{O}}\rangle$=$\langle \Psi |\hat{\mathcal{O}}|\Psi \rangle$. The consecutive terms in the 
above expression are the mass-velocity $\langle P_4 \rangle$, one-electron Darwin $\langle D_1 
\rangle$, two-electron Darwin $\langle D_2 \rangle$, and Breit $\langle B \rangle$ corrections, respectively. We assume 
that the value of the fine structure constant, $\alpha$, is $1/137.0359997$, as recommended by CODATA \cite{codata}.

Let us consider the values of the one-electron relativistic corrections, $\langle P_4 \rangle$ and $\langle D_1 
\rangle$. They can easily be obtained
within the STOs framework, since the corresponding one-electron integrals are fairly straightforward to compute. 
Integrals including the one-electron Dirac delta distribution reduce to the values of STOs at a given point of space 
which is elementary. Integrals including $\nabla^4$ operator reduce to combinations of the ordinary overlap integrals 
over STOs. General subroutines for calculation of the aforementioned integrals are now a part of our STOs integral 
package. Note, that $\langle P_4 \rangle$ and $\langle D_1 \rangle$ corrections (called also collectively the 
Cowan-Griffin contribution \cite{cowan76}) are very sensitive to the quality of the wavefunction in the vicinity of the 
nuclei. Therefore, their evaluation by using the STOs basis set is supposed to be particularly advantageous.

In Table \ref{relatom} we present values of the one-electron relativistc corrections, calculated with S-extended STOs 
basis sets. The results are compared with the values reported recently \cite{pachucki04a} which are considered 
``exact'' in the present context. Remarkably, in the biggest basis set, ATC-ETCC-6+S, the relative error of our values 
compared with the accurate ones is only $\approx$0.03\% and $\approx$0.003\% for $\langle P_4 \rangle$ and 
$\langle D_1 \rangle$, respectively. 
Moreover, even in the smallest basis set, ATC-ETCC-2+S, these errors increase to only about 0.1\% and 0.005\%. We found 
that it is 
impossible to reach a similar level of accuracy with the available (decontracted) GTOs basis sets, and typically the 
resulting error is (at least) an order of magnitude larger. 

It is also interesting to perform extrapolations of the values of one-electron relativistic corrections towards 
CBS. We found empirically that the following formulae provide the best fit
\begin{align}
\label{st1}
A + \frac{B}{(L+1)^2} \;\;\;\mbox{for $P_4$},\\
\label{st2}
A + \frac{B}{(L+1)^4} \;\;\;\mbox{for $D_1$}.
\end{align}
 Results of the extrapolations from $L$ = 3,4,5,6 are presented in Table 
\ref{relatom}. The extrapolation reduces the error of the mass-velocity correction to 0.02\%, but increases it 
insignificantly for the one-electron Darwin correction.

\section{Beryllium dimer}
\label{sec:dimer}

\begin{table*}[ht]
\caption{Results of the four-electron valence FCI calculations for the beryllium dimer. $N_b$ denotes the number of 
basis set functions, $N_{SD}$ is the dimension of the Hamiltonian matrix in $A_g$ symmetry, $E_{HF}$ is 
the Hartree-Fock energy, $E_c$ is the correlation energy at FCI level, CP is the counterpoise correction (for 
BSSE) to the interaction energy and $D_e$ is the calculated CP-corrected FCI interaction 
energy. The values in the last row are the extrapolated CBS values (see the main text for the discussion). All values 
are given in the atomic units unless stated otherwise.}
\begin{ruledtabular}
\begin{tabular}{c|rrcccc}
\label{fci}
basis set & $N_b$ & $N_{SD}$ & $E_{HF}$ & $E_c$ & CP / $\mbox{cm}^{-1}$ & $D_e$ / $\mbox{cm}^{-1}$ \\
\hline\\[-2.1ex]
A-ETCC-2 & 54  & 237 548 & $-$29.1339418 & $-$0.1046873 & 12.5 & 273.8 \\[0.6ex]
A-ETCC-3 & 100 & 2 895 037 & $-$29.1341621 & $-$0.1070574 & 8.3  & 710.6 \\[0.6ex]
A-ETCC-4 & 168 & 23 685 257 & $-$29.1341745 & $-$0.1076392 & 4.1  & 802.9 \\[0.6ex]
A-ETCC-5 & 260 & 138 002 229 & $-$29.1341751 & $-$0.1078505 & 2.6  & 833.8 \\[0.6ex]
A-ETCC-6 & 384 & 663 593 429 & $-$29.1341754 & $-$0.1079423 & 1.8  & 845.7 \\[0.6ex]
\hline\\[-2.1ex]
CBS & $\infty$ & $\infty$ & $-$29.1341754 & $-$0.1080695 & 0.0 & 864.9 $\pm$ 1.7 \\
\end{tabular}
\end{ruledtabular}
\end{table*}

\subsection{Four-electron (valence) contribution}
\label{subsec:val}

From earlier studies of the beryllium dimer, it is well-known that a major contribution to the interaction energy 
comes from the correlations between valence electrons.
Freezing both $1s^2$ atomic orbitals makes the dimer effectively a four-electron system which can be 
successfully treated with FCI method in large basis sets. We performed the frozen-core FCI calculations in basis sets 
A-ETCC-L with $L$ = 2,...,6. The Abelian group, $D_{2h}$, was used in computations.
We believe these are the biggest valence FCI calculations ever performed for this system in terms of the
number of configurations included in construction of the Hamiltonian matrix. The results of the calculations are 
included in Table \ref{fci}. In all cases the counterpoise correction (CP) for the basis set superposition error 
(BSSE) was applied \cite{boys70}. It is clear, that the results are slowly convergent also with respect to the quality 
of the basis set. This is probably due to the fact that bonding significantly perturbs the atomic densities. The 
increment of the interaction energy between $L$ = 5 and $L$ = 6 basis sets is as large as 11.9 $\mbox{cm}^{-1}$, 
suggesting that the CBS value is still significantly below the $L$ = 6 value.

Because of this observation it is necessary to perform some kind of extrapolation towards the CBS. The correlation 
energy alone was the subject of the extrapolation, separately for the atom and for the dimer. We used the formula
(\ref{extra}) which was previously used successfully for the atomic calculations. We also observe that in the largest
basis set, the Hartree-Fock (HF) results are already converged at least to eight significant digits. It is therefore
unnecessary to extrapolate the HF results and simply the value obtained in L=6 basis was taken as the CBS result. 

\begin{figure}[h]
\caption{Quality of the extrapolation towards the complete basis set for the beryllium dimer using results from basis 
sets A-ETCC-L with $L$ = 2,...,6 based on the theoretical expression (\ref{extra}). The dashed line denotes the 
estimated limit.}
\label{be2ex}
\includegraphics[scale=1.0]{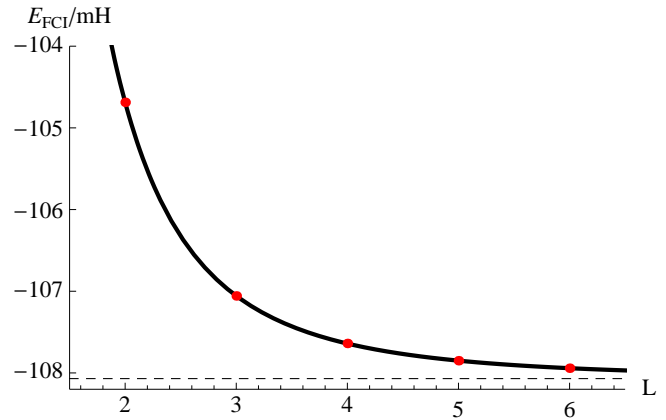}
\end{figure}

Note, that the CBS increment found in the extrapolation of the correlation energy is quite substantial, and crucial for 
the final results. It amounts to as much as nearly 20 $\mbox{cm}^{-1}$ in the interaction energy. Thus, it is necessary 
to additionally verify the reliability of the extrapolation. To do so, we first performed the extrapolation from 
$L$ = 2,3,4,5 basis sets in order to estimate the L = 6 value. The extrapolated L = 6 value gives the interaction 
energy 
equal
to 847.4 $\mbox{cm}^{-1}$ whereas the corresponding true calculated result is 845.7 $\mbox{cm}^{-1}$. The difference, 
amounting to 1.7 $\mbox{cm}^{-1}$, is assumed to be also the error of the CBS extrapolation from $L$ = 2,3,4,5,6. 
Quality of the extrapolation for the dimer is illustrated at Figure \ref{be2ex}. A quite similar excellent fit was 
obtained for the atomic calculations. Finally, our best estimate for the valence contribution to the interaction 
energy is 864.9 $\pm$ 1.7 $\mbox{cm}^{-1}$. Note, that this error estimation is a conservative one because 
extrapolation from a larger number of points can be expected to be more reliable. Additionally, the increment in the 
interaction 
energy between $L$ = 4 and $L$ = 5 basis set is significantly larger than between $L$ = 5 and $L$ = 6 or between $L$ = 
6 and the estimated CBS. Therefore, it is possible that our extrapolated result is more accurate then we assume here.

Our final result, namely 864.9 $\pm$ 1.7 $\mbox{cm}^{-1}$, is in line with recent findings of other authors. Patkowski 
\emph{et al.} \cite{patkowski07} found 857 $\pm$ 12 $\mbox{cm}^{-1}$, if we follow their method of error estimation, 
and Martin \cite{martin99} gives 872 $\pm$ 15 $\mbox{cm}^{-1}$. Present result lies well within the error bounds 
obtained in these works. A slight discrepancy is found between our result and the value recently reported by 
Evangelisti and co-workers \cite{helal13} who give 850.4 $\mbox{cm}^{-1}$ without any error estimation. We believe that 
this result is inaccurate, mainly because lack of the diffuse functions in their GTO basis set. Notably, our error 
bounds, which are conservative anyway, are an order of magnitude smaller than those obtained in the aforementioned 
works.

\subsection{Core-core and core-valence contribution}
\label{subsec:cccv}

\begin{table*}[ht]
\caption{Core-core and core-valence contribution to the interaction energy computed at various levels of theory. All 
values are given in $\mbox{cm}^{-1}$. Extrapolations are performed according to the formula (\ref{extra}), for the atom
and 
dimer separately, using the counterpoise-corrected data.}
\begin{ruledtabular}
\begin{tabular}{ccccccc}
\label{core}
basis set & $N_b$ & CCSD & CCSDT & CCSD(T) & MP2 & MP4 \\
\hline\\[-2.1ex]
ATC-ETCC-2 & 62  & 28.2 & 31.5 & 39.0 &-34.0 & 28.9 \\[0.6ex]
ATC-ETCC-3 & 126 & 50.4 & 56.7 & 61.2 & 57.2 & 56.4 \\[0.6ex]
ATC-ETCC-4 & 224 & 55.7 & 63.9 & 66.4 & 63.6 & 63.5 \\[0.6ex]
ATC-ETCC-5 & 364 & 57.4 &      & 67.7 & 65.7 & 65.7 \\[0.6ex]
\hline\\[-2.1ex]
CBS & $\infty$   & 59.3 & 69.6 & 69.5 & 67.8 & 68.4 \\
\end{tabular}
\end{ruledtabular}
\end{table*}

The second step in our calculations is a reliable determination of the core-core and core-valence contribution to the 
interaction energy. This task, however, is far from being trivial. A brief inspection of values available in the 
literature reveals that estimations from 65 $\mbox{cm}^{-1}$ \cite{helal13} to as large as 89 $\mbox{cm}^{-1}$ 
\cite{roeggen05} were obtained. Because of the fulfilment of the nuclear cusp condition, the STOs basis used in
the present work can be expected to be more suitable for the description of core region than the GTOs used thus far.

Our preliminary study suggests that the CCSDT model is a particularly good method for the estimation of the inner-shell 
contribution. The effect of connected quadruple excitations was found to be very small in this case. In fact, the 
effect of quadruples can be highly overestimated in small basis sets but quickly diminishes when the basis set is 
enlarged. We found this particular behaviour in virtually any approximate quadruples method that was available to us. 
Therefore, we can conclude that CCSDT method in the CBS limit would probably give the core-core and core-valence 
contribution accurate to within few parts in $\mbox{cm}^{-1}$. A similar observation was also made implicitly by Martin 
\cite{martin99}.

Unfortunately, we are able to perform all-electron CCSDT calculation only in ATC-ETCC-L basis sets with $L$ = 2,3,4. 
The results are 31.5 $\mbox{cm}^{-1}$, 56.7 $\mbox{cm}^{-1}$ and 63.9 $\mbox{cm}^{-1}$, respectively. 
CBS extrapolation from these values can be performed by using the formula (\ref{extra}), giving 69.6 $\mbox{cm}^{-1}$.
However, this three-point extrapolation is not particularly trustworthy since CBS increment is rather large and no 
reliable error estimation can be given. Thus, we must seek some approximate method, with smaller computational costs, 
giving results comparable to CCSDT in the CBS limit. 

\begin{figure}[h]
\caption{Contribution of the inner-shell effects to the interaction energy, denoted shortly $D_e^{core}$, calculated 
by using ATC-ETCC-L basis sets. Black dots are the CCSD(T) results and the black line is the CCSD(T)/CBS extrapolation 
curve. Analogously, red dots are the MP4 results and the red line is the corresponding CBS extrapolation. Blue squares 
are the available CCSDT results, for $L$ = 2,3,4.}
\label{be2cc}
\includegraphics[scale=1.0]{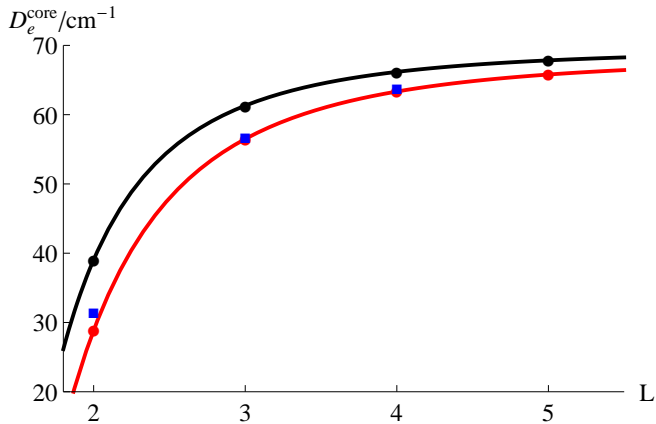}
\end{figure}

In Table \ref{core} we show inner-shell contributions to the interaction energy computed at various levels of theory.
CCSD, CCSD(T) and MP2 calculations were performed with \textsc{Gamess} package while CCSDT and MP4 energies were
evaluated with help of the \textsc{AcesII} program \cite{aces2}.
All values in this table were obtained by subtracting the interaction energy obtained with 
the frozen core approximation from the corresponding all-electron values. Let us compare the results of MP4 and CCSD(T) 
with the complete CCSDT model. One sees that MP4 method slightly underestimates the inner-shell contribution compared 
to CCSDT while CCSD(T) model overestimates it significantly, especially in smaller basis sets. Note additionally, that 
MP4 and CCSD(T) results strictly bracket the CCSDT values, as illustrated at Figure \ref{be2cc}. If we assume that this 
behaviour holds further then the CBS limit of the CCSDT method should lie between the corresponding limits of MP4 and 
CCSD(T). Fortunately, the CBS limit is 68.4 $\mbox{cm}^{-1}$ and 69.5 $\mbox{cm}^{-1}$ for MP4 and CCSD(T), 
respectively. The exact result probably lies between these values so as the final result we take the average of the two 
and estimate the error as a half of the difference between them. This gives the final value of the core-core and 
core-valence contributions to the interaction energy equal to 69.0 $\pm$ 0.6 $\mbox{cm}^{-1}$. The small effect of the 
connected quadruples contribution is probably already incorporated in the error estimation.

Note, that the final value determined by us is significantly smaller than some of the estimations given in the 
literature. For instance, Martin gives 76.2 $\mbox{cm}^{-1}$ \cite{martin99} while Patkowski \emph{et al.} 
\cite{patkowski07} reports as much as 85$\,\pm\,$5 $\mbox{cm}^{-1}$. We believe that these discrepancies are mainly due 
to defects in the GTOs basis sets used by authors. In fact, when GTOs basis sets are not designed very carefully in the 
core region, the inner-shell correlation effects can be significantly overestimated. Naturally, STOs are much more 
appropriate in this respect which is one of their noteworthy advantages.

\subsection{Relativistic, QED and adiabatic corrections}
\label{subsec:rel}

\begin{table*}[ht]
\caption{Mass-velocity, $\langle P_4 \rangle$, and one-electron Darwin, $\langle D_1 \rangle$, corrections for the 
beryllium dimer calculated at the 
CCSD and FCI levels of theory. The factor of $\alpha^2$ is not included. All values are given in the atomic units.}
\label{relbe2}
\begin{ruledtabular}
\begin{tabular}{ccccccc}
 & \multicolumn{2}{c}{all-electron CCSD} & \multicolumn{2}{c}{frozen-core CCSD} & \multicolumn{2}{c}{frozen-core FCI} 
\\[0.6ex]
\hline \\[-2.2ex]
basis set  & $\langle P_4 \rangle$ & $\langle D_1 \rangle$ & $\langle P_4 \rangle$ & $\langle D_1 \rangle$ & $\langle 
P_4 \rangle$ & $\langle D_1 \rangle$ \\[0.6ex]
\hline \\[-2.2ex]
ATC-ETCC2+S & $-$539.847 891 & 443.692 152 & $-$537.394 631 & 443.278 203 & $-$537.133 303 & 443.083 762 \\[0.6ex]
ATC-ETCC3+S & $-$539.971 064 & 443.675 656 & $-$537.333 536 & 443.241 928 & $-$537.036 087 & 443.021 849 \\[0.6ex]
ATC-ETCC4+S & $-$540.030 590 & 443.664 899 & $-$537.317 658 & 443.233 044 & $-$537.014 183 & 443.008 427 \\[0.6ex]
ATC-ETCC5+S & $-$540.073 538 & 443.665 227 & $-$537.310 464 & 443.229 426 & $-$537.004 508 & 443.003 144 \\[0.6ex]
\hline \\[-2.2ex]
CBS & $-$540.141 465 & 443.655 919 & $-$537.305 424 & 443.226 021 & $-$536.995 150 & 442.996 653 \\
\end{tabular}
\end{ruledtabular}
\end{table*}

\begin{table*}[ht]
\caption{Contributions to the interaction energy of the beryllium dimer from the mass-velocity, $\langle P_4 \rangle$, 
and one-electron Darwin, $\langle D_1 \rangle$, corrections calculated at the CCSD and FCI levels of theory. All 
results are given in 
cm$^{-1}$.}
\label{intrel}
\begin{ruledtabular}
\begin{tabular}{cccccccccc}
 & \multicolumn{3}{c}{all-electron CCSD} & \multicolumn{3}{c}{frozen-core CCSD} & \multicolumn{3}{c}{frozen-core FCI} \\
\hline \\[-2.2ex]
basis set & $D_e( P_4 )$ & $D_e( D_1 )$ & $\Sigma D_e$
& $D_e( P_4 )$ & $D_e( D_1 )$ & $\Sigma D_e$
& $D_e( P_4 )$ & $D_e( D_1 )$ & $\Sigma D_e$ \\
\hline \\[-2.2ex]
ATC-ETCC2+S & $-$12.40 & 9.28  & $-$3.12 & $-$11.87 & 8.87 & $-$3.00 & $-$14.93 & 11.14 & $-$3.78 \\[0.6ex]
ATC-ETCC3+S & $-$13.35 & 9.79  & $-$3.57 & $-$12.71 & 9.41 & $-$3.30 & $-$16.19 & 11.98 & $-$4.21 \\[0.6ex]
ATC-ETCC4+S & $-$13.63 & 10.08 & $-$3.54 & $-$12.94 & 9.56 & $-$3.38 & $-$16.49 & 12.18 & $-$4.30 \\[0.6ex]
ATC-ETCC5+S & $-$13.72 & 10.14 & $-$3.58 & $-$13.03 & 9.61 & $-$3.42 & $-$16.60 & 12.25 & $-$4.35 \\[0.6ex]
\hline \\[-2.2ex]
CBS         & $-$14.26 & 10.44 & $-$3.81 & $-$13.10 & 9.67 & $-$3.44 & $-$16.73 & 12.34 & $-$4.39 \\
\end{tabular}
\end{ruledtabular}
\end{table*}

One-electron relativistic corrections were evaluated by using the S-extended basis sets, described in Section
\ref{sec:pre}. The results are presented in Table \ref{relbe2}. Calculations of the one-electron expectation values, at 
the all-electron and frozen-core CCSD level of theory, were performed by using $\Lambda$ operator technique 
\cite{fitzgerald86,salter87,salter89,wloch05} implemented by default in \textsc{Gamess} package. Relaxation of the 
Hartree-Fock orbitals is neglected in CCSD calculations. FCI calculations were done using our own program and the 
expectation values are straightforward to evaluate by using the FCI wavefunctions. 

Extrapolations are carried out by using the empirical formula (\ref{st1}) for \emph{both} $\langle D_1 
\rangle$ and $\langle P_4 \rangle$. Our strategy for evaluation of the contribution to the interaction energy from the 
Cowan-Griffin approximation \cite{cowan76} is as follows. We use the valence FCI values corrected for the core-core and 
core-valence effects, as a difference between all-electron and frozen-core CCSD results. It was found previously that 
CCSD method behaves reasonably for the inner-shell correlations (see Table \ref{core}) and this accuracy is sufficient 
for the 
present purposes. In Table \ref{intrel} we present contributions to the interaction energy from $\langle D_1 \rangle$ 
and $\langle P_4 \rangle$ corrections, calculated at this level of theory. The core-core and 
core-valence CCSD effect is estimated to be $-$0.4 $\mbox{cm}^{-1}$, while the pure valence FCI contribution is $-$4.4 
$\mbox{cm}^{-1}$.  By summing both corrections we obtain $-$4.8 $\pm$ 0.2 $\mbox{cm}^{-1}$ for the final contribution 
to the interaction energy coming from the one-electron 
relativistic corrections. The error is simply taken as the (rounded up) value of the corresponding CBS increment. The 
obtained value is in a moderate agreement with the values given by Patkowski \emph{et al.} \cite{patkowski07}, $-$4.1 
$\mbox{cm}^{-1}$, Martin \cite{martin99}, $-$4.0 $\mbox{cm}^{-1}$, and Gdanitz \cite{gdanitz99}, $-$5.2
$\mbox{cm}^{-1}$. 
However, as far as we can tell, these values are not extrapolated and the authors report no respective error bars of 
their result. We believe that our final values are much more accurate due to the fact 
that STOs basis sets were used throughout.

\begin{table*}[ht]
\caption{Contributions to the interaction energy of the beryllium dimer from the two-electron Darwin, $\langle D_2 
\rangle$, and Breit, $\langle B \rangle$, corrections calculated at the CCSD(T) and FCI levels of theory within GTOs 
basis sets. All results are given in cm$^{-1}$.}
\label{intrel2}
\begin{ruledtabular}
\begin{tabular}{cccccccccc}
 & \multicolumn{3}{c}{all-electron CCSD(T)} & \multicolumn{3}{c}{frozen-core CCSD(T)} & \multicolumn{3}{c}{frozen-core 
FCI} \\
\hline \\[-2.2ex]
basis set & $D_e( D_2 )$ & $D_e( B )$ & $\Sigma D_e$
& $D_e( D_2 )$ & $D_e( B )$ & $\Sigma D_e$
& $D_e( D_2 )$ & $D_e( B )$ & $\Sigma D_e$ \\
\hline \\[-2.2ex]
aug-cc-pCVDZ & 0.38 & $-$0.82 & $-$0.44 & 0.41 & $-$0.73 & $-$0.32 & 0.42 & $-$0.76 & $-$0.34 \\[0.6ex]
aug-cc-pCVTZ & 0.42 & $-$0.89 & $-$0.47 & 0.46 & $-$0.77 & $-$0.32 & 0.46 & $-$0.80 & $-$0.34 \\[0.6ex]
aug-cc-pCVQZ & 0.43 & $-$0.90 & $-$0.47 & 0.47 & $-$0.79 & $-$0.31 & 0.48 & $-$0.82 & $-$0.34 \\[0.6ex]
aug-cc-pCV5Z & 0.44 & $-$0.91 & $-$0.47 & 0.48 & $-$0.79 & $-$0.31 & 0.48 & $-$0.82 & $-$0.34 \\[0.6ex]
\end{tabular}
\end{ruledtabular}
\end{table*}

Let us now focus on the two-electron relativistic corrections - two-electron Darwin, $\langle D_2 \rangle$, and Breit, 
$\langle B \rangle$, contributions. Evaluation of the latter correction within the STOs basis set is not feasible at
present. This is mostly due to the fact the matrix elements of the Breit term, Eq. (\ref{breit}), are 
extremely difficult to compute with the exponential functions. As far as we know, the only accurate molecular 
calculations of the Breit term within the exponential basis set were performed by Ko\l os and Wolniewicz 
\cite{kolos64,wolniewicz95} for various electronic states of H$_2$. 

Because of these difficulties, we calculated $\langle D_2 \rangle$ and $\langle B \rangle$ in GTOs
basis sets. It will be shown that contributions of the two-electron relativistic 
corrections are small and GTOs basis sets are sufficient to meet the prescribed accuracy requirements.

For calculations of the two-electron relativistic corrections we used modified aug-cc-pCVXZ series of GTOs basis sets 
\cite{dunning89,dunning92,dunning93,dunning94,dunning95,dunning96,dunning99,dunning01,dunning11}. To improve the 
quality of the wavefunction the standard set of $1s$ GTOs orbitals was replaced by a new one comprising 23 $1s$ 
functions. This set was obtained by minimising the Hartree-Fock energy of the beryllium atom. Apart from that, the 
original $1s$ diffuse functions from the initial aug-cc-pCVXZ basis sets were kept. We also decontracted the 
$2p$ polarisation functions and removed the redundant orbitals. Higher angular momentum shells were neither modified 
nor decontracted.

\textsc{Dalton} program package \cite{dalton} was used for CCSD(T) calculations and our own program for the valence FCI 
calculations. In Table \ref{intrel2} we show contributions of $\langle D_2 \rangle$ and $\langle B \rangle$ to the 
interaction energy computed at three different levels of theory - all-electron and frozen-core CCSD(T), and frozen-core 
FCI. It is not necessary to perform CBS extrapolations since the contributions to the interaction energy are converged 
to about 0.01$-$0.02 $\mbox{cm}^{-1}$ already in the biggest basis set. We take the frozen-core FCI contribution as our 
result and additionally correct it for the inner-shell effects as a difference between the all-electron and frozen-core 
CCSD(T) values. In this way, we obtain the contribution to the interaction energy from the two-electron relativistic 
correction equal to $-$0.5 $\mbox{cm}^{-1}$. The error can be estimated to be much below 0.1 $\mbox{cm}^{-1}$ by 
observing the convergence pattern in the available basis sets. Unfortunately, we are not aware of any available 
literature values that we could compare with.

By summing the computed one- and two-electron relativistic contributions, we find that $\alpha^2$ effects decrease the
interaction energy by 5.3 $\pm$ 0.2 $\mbox{cm}^{-1}$. This contribution is quite sizable and definitely needs to be 
included to obtain a spectroscopically accurate potential energy curve for the beryllium dimer.

Let us now pass to the leading-order QED contribution. Theoretically, this effect should be by a factor $\alpha$ 
smaller than the Breit-Pauli contribution and thus entirely negligible within the present accuracy requirements.
However, it turns out that among the relativistic contributions to the interaction energy there is a significant
cancellation between $\langle P_4 \rangle$ and $\langle D_1 \rangle$ terms, so that the result is order of magnitude 
smaller than the net values of separate terms. Therefore, the leading QED corrections may still contribute to the 
interaction energy significantly. In fact, this situation was previously encountered in calculations for the dihydrogen 
\cite{piszczatowski09} and the helium dimer \cite{cencek12}. This suggests that whenever the $\alpha^2$ relativistic 
corrections are included in accurate calculations for light systems, the leading-order QED contributions should also be 
at least estimated.

The leading QED correction (of the order $\alpha^3$ and $\alpha^3 \ln \alpha$) to the electronic energy of a
molecular singlet state takes the form \cite{araki57,sucher58}
\begin{align}
\label{qed}
\begin{split}
E^{(3)} &= \frac{8\alpha}{3\pi}\left(\frac{19}{30}-2\ln \alpha - \ln k_0\right)\langle D_1 \rangle \\
&+ \frac{\alpha}{\pi} \left(\frac{164}{15}+\frac{14}{3}\ln \alpha\right)\langle D_2 \rangle + \langle 
H_{AS} \rangle,
\end{split}
\end{align}
where $\ln k_0$ is the so-called Bethe logarithm \cite{bethe75,schwartz61}, $\langle D_1 \rangle$ and $\langle D_2 
\rangle$ are the values of the one- and two-electron Darwin corrections (including the factor of $\alpha^2$). The term 
$\langle H_{AS} \rangle$ is the Araki-Sucher contribution, given by the following expectation value
\begin{align}
\langle H_{AS} \rangle = -\frac{7\alpha^3}{6\pi} \langle \sum_{i>j} \hat{P}\left(r_{ij}^{-3}\right)\rangle,
\end{align}
and $\hat{P}\left(r_{ij}^{-3}\right)$ denotes the regularised $r_{ij}^{-3}$ distribution,
\begin{align}
\langle\hat{P}\left(r_{ij}^{-3}\right)\rangle = \lim_{a\rightarrow 0}\langle
\theta(r_{ij}-a)r_{ij}^{-3}+4\pi (\gamma_E+\ln a)\delta(\textbf{r}_{ij}) \rangle,
\end{align}
where $\gamma_E$ is the Euler-Mascheroni constant.
It is well know that computation of the Bethe logarithm and Araki-Sucher terms is extremely difficult and has never
been attempted for any molecular system apart from the dihydrogen \cite{piszczatowski09} and the helium dimer 
\cite{cencek12}. Therefore, we have to adopt some approximate strategy for determination of $E^{(3)}$. Fortunately, 
except at very large $R$, the Araki-Sucher term is small compared to the overall leading-order QED correction and thus 
can be neglected. The Bethe logarithm, on the other hand, was found to vary insignificantly as the function of 
$R$, when $R$ is moderate (or large), for the helium dimer and dihydrogen. Therefore, the asymptotic (atomic) value of 
the Bethe logarithm can be adopted.

A very accurate value of $\ln k_0$ for the beryllium atom has been given recently by Pachucki and Komasa 
\cite{pachucki04a}, $\ln k_0 
= 5.75034$. We use the extrapolated values of $\langle D_1 \rangle$ and $\langle D_2 \rangle$, equal to $0.023613$, 
$0.000522$ for the dimer, and $0.011836$, $0.000262$ for the monomer, respectively. With these assumptions, 
contribution of the lowest-order QED effects to the interaction energy of the beryllium dimer is calculated to be 
0.37 $\mbox{cm}^{-1}$. This value is an order of magnitude smaller than the relativistic corrections, as expected. 
However, their omission would significantly increase the total error of our theoretical predictions. It is difficult 
to estimate strictly what is the effect of the adopted approximations on the value of QED contribution to the 
interaction energy. For the dihydrogen molecule, exactly the same approximations introduce an error slightly less than 
10\%, basing on the results presented in Ref. \cite{piszczatowski09}. Therefore, we can assume very conservatively that 
error of the 
present calculations is at most 20\%. This finally gives us estimation of the leading-order QED contribution to the 
interaction energy equal to 0.4$\,\pm\,$0.1 $\mbox{cm}^{-1}$.

We also check the next higher-order QED contribution. It is well 
known from the calculations on the helium atom \cite{pachucki06a,pachucki06b}, that the $\alpha^4$ effects are 
dominated by the one-loop term \cite{eides01} given by
\begin{align}
E^{(4)}_{\rm one-loop} = 16 \alpha^2 \left( \frac{427}{192}-\ln 2\right)\langle D_1 \rangle,
\end{align}
in the case of the beryllium atom (or dimer).
The above quantity is a scaled one-electron Darwin correction and thus can be easily computed. We found that 
the contribution to the interaction energy of the one-loop term to be approximately $0.017$ $\mbox{cm}^{-1}$, which is 
well below 0.1 $\mbox{cm}^{-1}$. Therefore, as anticipated, the higher-order QED contributions can safely be neglected 
within the present accuracy requirements. This additionally gives a verification that the QED perturbative 
series converges rapidly for the beryllium dimer.

The remaining missing part of the theory that has to be investigated is the finite nuclear mass \emph{i.e.} 
the adiabatic correction. We calculated this correction with help of the \textsc{Cfour} \cite{cfour} and \textsc{mrcc} 
\cite{mrcc1,mrcc2} program packages at both all-electron and frozen-core CCSD and CCSDT levels of theory 
\cite{gauss06}. GTOs basis set which were previously used for 
computation of the two-electron relativistic corrections were utilised. In all cases we found that the contribution to 
the interaction energy from the adiabatic correction was significantly below 0.1 $\mbox{cm}^{-1}$. In fact, the net 
values of the adiabatic correction for both atom and dimer were large, but they cancelled out almost to zero. This is 
probably due to the fact that the adiabatic correction contribution to the interaction energy as a function of the 
internuclear distance, $R$, crossed zero near the value of $R$ adopted by us (close to the minimum). A similar 
situation was found in the case of the helium dimer \cite{cencek12}. Our observation is additionally verified by 
calculations of Koput \cite{koput11} who found that contribution of the adiabatic correction to the interaction energy 
varies by only 2 $\mbox{cm}^{-1}$ along the whole potential energy curve. As a result, we assume that the contribution 
to the interaction energy coming from the adiabatic effects is equal to zero. We estimate that the error of this result 
is at most 0.1 $\mbox{cm}^{-1}$.

\subsection{Total interaction energy}
\label{subsec:finalbe2}

\begin{table}[hb]
\caption{Final error budget of the calculations for the ground state $\left(^1\Sigma_g^+\right)$ of the beryllium dimer 
obtained in this 
work. All values are given in cm$^{-1}$.}
\label{bugdet}
\begin{ruledtabular}
\begin{tabular}{lr}
 & contribution to $D_e$ \\[0.6ex]
\hline \\[-2.2ex]
valence correlations & $+$864.9$\,\pm\,$1.7 \\[0.6ex]
inner-shell correlations & $+$69.0$\,\pm\,$0.6 \\[0.6ex]
relativistic ($\alpha^2$) effects & $-$5.3$\,\pm\,$0.2 \\[0.6ex]
leading-order ($\alpha^3$) QED effects & $+$0.4$\,\pm\,$0.1 \\[0.6ex]
adiabatic correction & $+$0.0$\,\pm\,$0.1 \\[0.6ex]
\hline \\[-2.2ex]
total & $+$929.0$\,\pm\,$1.9 \\[0.6ex]
\hline \\[-2.2ex]
experiment & $+$929.7$\,\pm\,$2.0 \\
\end{tabular}
\end{ruledtabular}
\end{table}

\begin{table*}[ht]
\caption{Results of the selected theoretical predictions for the ground state of the beryllium dimer published since 
late 90'. All values are given in cm$^{-1}$ and error bars are shown if estimated originally. Relativistic corrections 
are included if calculated. AE and FC denote all-electron and frozen-core, respectively. A majority of the acronyms 
appearing below is explained in the main text, apart from: ACPF - averaged coupled-pair functional, CC3 - coupled 
cluster model with an approximate treatment of triple excitations, CAS - complete active space, MR-CISD+Q - 
multireference configuration interaction with single and double excitations, Q denotes a specific Davidson-type 
correction for lack of size extensivity.}
\label{theory}
\begin{ruledtabular}
\begin{tabular}{clll}
 year & method & $D_e$ & reference \\[0.6ex]
\hline \\[-2.2ex]
1999 & FC CCSD(T)+FCI/CBS and AE CAS-ACPF & 944$\,\pm\,$25 & Martin \cite{martin99} \\[0.6ex]
1999 & CAS $r_{12}$-MR-ACPF/GTO($19s11p6d4f3g2h$) & 898$\,\pm\,$8 & Gdanitz \cite{gdanitz99} \\[0.6ex]
2000 & CC3+FCI/d-aug-cc-pVQZ & 885 & Pecul \emph{et al.} \cite{pecul00} \\[0.6ex]
2005 & EXRHF/GTO($23s10p8d6f3g2h$) & 945$\,\pm\,$15 & R\o{}ggen and Veseth \cite{roeggen05} \\[0.6ex]
2007 & AE CCSD(T)/CBS and FC FCI/CBS & 938$\,\pm\,$15 & Patkowski \emph{et al.} \cite{patkowski07} \\[0.6ex]
2007 & variational Monte Carlo and fixed-node diffusion Monte Carlo & 829$\,\pm\,$64 & Harkless and Irikura 
\cite{harkless06} \\[0.6ex]
2010 & FC FCI/CBS and AE MR-CISD+Q & 912 & Schmidt \emph{et al.} \cite{schmidt10} \\[0.6ex]
2010 & AE MRCI/CBS & 818 & Mitin \cite{mitin11} \\[0.6ex]
2011 & AE CCSD(T)/CBS and FC FCI/CBS & 935$\,\pm\,$10 & Koput \cite{koput11} \\[0.6ex]
2013 & FC FCI/CBS and AE CCSD(T)/cc-pV6Z & 927.4$\,\pm\,$12 & Evangelisti \emph{et al.} \cite{helal13} \\[0.6ex]
2014 & density matrix renormalisation group (DMRG) & 931.2 & Sharma \emph{et al.} \cite{sharma14} \\[0.6ex]
present & FC FCI/CBS and AE CCSD(T)/MP4/CBS & 929.0$\,\pm\,$1.9 & --- \\[0.6ex]
\end{tabular}
\end{ruledtabular}
\end{table*}

All contributions to the interaction energy of the beryllium dimer computed in this work are listed in Table 
\ref{bugdet}. By summing all contributions we obtain the value 929.0 $\mbox{cm}^{-1}$ which is the main result of our 
study. The overall error of the calculations is estimated by summing squares of all fractional errors (1.7, 0.6, 
0.2, 0.1, 0.1 $\mbox{cm}^{-1}$) and taking the square root, which gives 1.9 $\mbox{cm}^{-1}$ (rounded up) or 0.2\%. The 
total result, 929.0$\,\pm\,$1.9 $\mbox{cm}^{-1}$, is in a very good agreement with the latest experimental value, 
929.7$\,\pm\,$2.0 $\mbox{cm}^{-1}$, reported by Merritt \emph{et al.} \cite{merritt09}. In fact, the present result 
lies within the error bars of the empirical value and \emph{vice versa}. 

Let us also comment on the timings of the present calculations. It is true that any gain connected with 
the use of STOs can easily diminish if computation of the STOs two-electron integral files becomes overwhelmingly time 
consuming, up to a point when it is more expensive than evaluation of the molecular energy. There is such a risk, 
because STOs integral algorithms are inherently more complicated and demanding than their GTOs counterparts. In fact, 
we found that calculation of the STOs integrals is one or two orders of magnitude more expensive than in the case of 
GTOs, with the same size of the basis set. This sounds daunting but the actual situation is more complex. For 
instance, in the largest basis sets used in this work, the calculation of the GTOs two-electron integrals is a matter 
of several minutes while in STOs it takes up to few hours. However, full CI or high-level CC calculations typically 
take several days to converge. Therefore, calculation of the integral files constitutes a small fraction of the total 
timing and does not pose any practical bottleneck. This is clearly a consequence of relatively low scaling ($N^4$)
of the calculations of the integral files, as compared with high-level CC of FCI methods.

It is also worthy to compare our results with the latest theoretical values predicted by other authors. In Table 
\ref{theory} we collected most of the theoretical results published in the late 90' and since then. An extensive 
bibliography of calculations published prior to this date can be found in Refs. \cite{roeggen05} and 
\cite{patkowski07}. Probably the most reliable 
calculations given thus far for the beryllium dimer are those of Patkowski \emph{et al.} \cite{patkowski07}, giving 
938$\,\pm\,$15 $\mbox{cm}^{-1}$, and Koput \cite{koput11}, 935$\,\pm\,$10 $\mbox{cm}^{-1}$. Our result is slightly 
lower but it lies within the error bars estimated by authors. Remarkably, the error predicted by us is by an order of 
magnitude smaller than in the previous works, despite our estimations were rather conservative.  Therefore, it seems 
that the theoretical values published thus far converge towards a value around 930 $\mbox{cm}^{-1}$, very close to the 
recent experimental result.

Apart from that, it is worthy to quote three semiempirical results obtained by ``morphing'' the 
theoretical potential energy curve in order to reproduce the experimentally measured vibrational levels 
\cite{patkowski09}. These values are 933.0, 933.2 and 934.6 $\mbox{cm}^{-1}$. It is difficult to estimate the error of 
these values but we feel that these semiempirical results are also consistent with our final value, 929.0$\,\pm\,$1.9 
$\mbox{cm}^{-1}$.

\section{Conclusions and outlook}
\label{sec:conclusion}
We have obtained a reliable value of the interaction energy for the beryllium dimer by using STOs 
basis sets combined with high-level quantum chemistry methods. The total error estimated by us, 1.9 $\mbox{cm}^{-1}$, 
is by an order of magnitude smaller than in the previous theoretical works. The most striking advantages of STOs, as 
compared with GTOs, are the reliability in estimation of the core-core and core-valence correlation effects, very 
solid quality of extrapolations towards CBS, and improved performance in calculation of the one-electron 
relativistic effects. It is clear that all of these features are essential for a spectroscopically accurate 
determination of the potential energy curves for diatomic systems. We have not found a situation when STOs perform 
worse than GTOs basis sets of the same size, at least among those available to us. Despite the fact that the evaluation
of the two-electron integrals in STOs basis is much more computationally intensive than in
the case of GTOs, we have never found it to be a practical bottleneck. An obvious disadvantage of STOs is the fact that
two-electron two-centre integrals which are required for calculation of the Breit $\alpha^2$ relativistic correction are
very difficult to compute and we needed to resort to GTOs to compute them.

It is also worthy to consider the direction of further 
advancements which can be taken. Let us recall the fact, that the ground state of the beryllium dimer is a very 
pathological and difficult system \emph{e.g.} the triple excitations are responsible for the bonding 
effects. In many different spectroscopically interesting diatomic systems the situation is not that difficult and the 
doubly excited determinants give the dominant contribution to the interaction energy. In such situations the explicitly 
correlated calculations \cite{hattig12,kong12} are an option, allowing for a much better saturation at the MP2, CCD or 
CCSD levels of theory. The F12 theory of explicitly correlated calculations is now well established \cite{may04} but to 
apply STOs in such computations several issues of both technical and theoretical nature need to be resolved. For 
instance, for GTOs 
calculations the exponential correlation factor of Ten-no \cite{tenno04a,tenno04b} is nowadays routinely used. In the 
case of STOs basis sets 
this choice is not feasible at present, due to an extremely complicated theory of evaluation of the resulting molecular 
two-electron integrals \cite{lesiuk12,pachucki12b}. Therefore, a different correlation factor has to be adapted. Other 
problems such as quality and 
design of the auxiliary basis sets \cite{klopper02,valeev04} for the resolution of identity approximation also need to 
be addressed. Nonetheless, 
the work on combining STOs basis sets with explicitly correlated theories is in progress in our laboratory.

Let us suppose that the accuracy of calculation of the Born-Oppenheimer potential energy curves can be further improved 
by an order of magnitude, say, due to use of the explicitly correlated methods and other theoretical advancements. The 
dominant error would then come from inaccuracies in calculation of the relativistic effects, especially for heavier 
systems. If a perturbation theory, using the Breit-Pauli Hamiltonian, can be still applied then it is natural that 
two-electron relativistic effects should be calculated within the STOs basis sets. Therefore, sooner or later we shall 
face the problem of evaluation of the matrix elements of the orbit-orbit and spin-orbit operators with the exponential 
functions. For heavy atoms, where the perturbation theory breaks down, different approaches need to be considered 
such as Douglas-Kroll-Hess transformations \cite{hess86,hess89,hess92,hess02} or use of effective core potentials 
\cite{kahn76,dolg12}. Neither of the above methods can straightforwardly be combined with the STOs basis sets. 
Nonetheless, our preliminary studies showed that extensions in these directions are feasible.

We can conclude by noting that the present series of papers opens up a possibility for a significant increase of 
accuracy which can be routinely reached for the diatomic systems with \emph{ab-initio} methods.

\begin{acknowledgments}
This work was supported by the Polish Ministry of Science and Higher Education, grant NN204 182840. ML acknowledges the 
Polish Ministry of Science and Higher Education for the support through the project \textit{``Diamentowy Grant''}, 
number DI2011 012041. RM was supported by the Foundation for Polish Science through the \textit{``Mistrz''} program.
\end{acknowledgments}

\end{document}


\title{{\Large Supplemental Material.} \\[0.5em] \noindent 
Calculation of two-centre two-electron integrals over Slater-type orbitals revisited. \\III.
Case study of the beryllium dimer}

\author{\sc Micha\l\ Lesiuk}
\email{e-mail: lesiuk@tiger.chem.uw.edu.pl}
\author{\sc Micha\l\ Przybytek}
\affiliation{\sl Faculty of Chemistry, University of Warsaw, Pasteura 1, 02-093 Warsaw, Poland}
\author{\sc Monika Musia\l}
\affiliation{\sl Institute of Chemistry, University of Silesia, Szkolna 9, 40-006 Katowice, Poland}
\author{\sc Bogumi\l\ Jeziorski}
\author{\sc Robert Moszynski}
\affiliation{\sl Faculty of Chemistry, University of Warsaw, Pasteura 1, 02-093 Warsaw, Poland}
\date{\today}

\begin{abstract}
The present document serves as a supplemental material for the publication \emph{Calculation of two-centre
two-electron integrals over Slater-type orbitals revisited. III. Case study of the beryllium dimer}. It
contains data or derivations which are of some importance but are not necessary for an overall understanding of the
manuscript. However, the material presented here will be useful for a reader who wishes to repeat all of our derivations
in details. Researchers who wish to repeat some of the calculations independently may also benefit from these
additional data.
\end{abstract}

\maketitle

\section{Supplemental material for Paper III}

\setlength{\tabcolsep}{12pt}
\renewcommand{\arraystretch}{1.2}
\begin{table}[h]
\caption{Composition of STO basis sets for the beryllium atom. ETCC-2 basis sets contains all functions from the first
column. TC-ETCC-2 basis set contains additionally functions from the second column and ATC-ETCC-2 functions from the
third column. A-ETCC-2 contains functions from the first and third columns. The symbol $[k]$ denotes the powers of 10,
10$^k$.}
\begin{tabular}{cc|cc|cc}
\hline\hline
\multicolumn{2}{c|}{atomic valence} & 
\multicolumn{2}{c|}{tight core} & 
\multicolumn{2}{c}{diffuse} \\
\hline\\[-3.2ex]
1S & 8.12906491[$-$01] &
1S & 1.17156475[+01] &
1S & 7.16639837[$-$01] \\
1S & 1.15567845[+00] &
2P & 5.29709026[+00] &
2P & 1.05907592[+00] \\
1S & 1.64298439[+00] &
   &                &
3D & 1.16078021[+00] \\
1S & 2.33576885[+00] &
   &                &
   &                \\
1S & 3.32067433[+00] &
   &                &
   &                \\
1S & 4.72087725[+00] &
   &                &
   &                \\
1S & 6.71149285[+00] &
   &                &
   &                \\
2P & 1.25956795[+00] &
   &                &
   &                \\
2P & 1.34363972[+00] &
   &                &
   &                \\
3D & 1.22675519[+00] &
   &                &
   &                \\
\hline\hline
\end{tabular}
\end{table}

\begin{table}
\caption{Composition of STO basis sets for the beryllium atom. ETCC-3 basis sets contains all functions from the first
column. TC-ETCC-3 basis set contains additionally functions from the second column and ATC-ETCC-3 functions from the
third column. A-ETCC-3 contains functions from the first and third columns. The symbol $[k]$ denotes the powers of 10,
10$^k$.}
\begin{tabular}{cc|cc|cc}
\hline\hline
\multicolumn{2}{c|}{atomic valence} & 
\multicolumn{2}{c|}{tight core} & 
\multicolumn{2}{c}{diffuse} \\
\hline\\[-3.2ex]
1S & 8.07003147[$-$01] &
1S & 1.54935930[+01] &
1S & 5.78796614[$-$01] \\
1S & 1.17156475[+00] &
1S & 2.05467840[+01] &
2P & 1.01069074[+00] \\
1S & 1.70081613[+00] &
2P & 6.63663083[+00] &
3D & 1.13131315[+00] \\
1S & 2.46915546[+00] &
2P & 7.90324352[+00] &
4F & 1.09034661[+00] \\
1S & 3.58459012[+00] &
3D & 7.77891287[+00] &
   &                \\
1S & 5.20391953[+00] &
   &                &
   &                \\
1S & 7.55477686[+00] &
   &                &
   &                \\
1S & 1.09676280[+00] &
   &                &
   &                \\
2P & 1.07328789[+00] &
   &                &
   &                \\
2P & 1.77687814[+00] &
   &                &
   &                \\
2P & 2.94170463[+00] &
   &                &
   &                \\
3D & 1.30383348[+00] &
   &                &
   &                \\
3D & 2.13842350[+00] &
   &                &
   &                \\
4F & 1.64010475[+00] &
   &                &
   &                \\
\hline\hline
\end{tabular}
\end{table}

\begin{table}
\caption{Composition of STO basis sets for the beryllium atom. ETCC-4 basis sets contains all functions from the first
column. TC-ETCC-4 basis set contains additionally functions from the second column and ATC-ETCC-4 functions from the
third column. A-ETCC-4 contains functions from the first and third columns. The symbol $[k]$ denotes the powers of 10,
10$^k$.}
\begin{tabular}{cc|cc|cc}
\hline\hline
\multicolumn{2}{c|}{atomic valence} & 
\multicolumn{2}{c|}{tight core} & 
\multicolumn{2}{c}{diffuse} \\
\hline\\[-3.2ex]
1S & 5.56501242[$-$01] &
1S & 1.42014407[+01] &
1S & 4.33207693[$-$01] \\
1S & 8.08302238[$-$01] &
1S & 1.82716996[+01] &
2P & 6.71390454[$-$01] \\
1S & 1.17403603[+00] &
2P & 7.37173108[+00] &
3D & 7.58415000[$-$01] \\
1S & 1.70525397[+00] &
2P & 9.86210540[+00] &
4F & 8.96307000[$-$01] \\
1S & 2.47683292[+00] &
2P & 1.20127077[+01] &
5G & 1.02303000[+00] \\
1S & 3.59752943[+00] &
3D & 8.37590965[+00] &
   &                \\
1S & 5.22530925[+00] &
3D & 1.01997372[+01] &
   &                \\
1S & 7.58961318[+00] &
4F & 1.03564887[+01] &
   &                \\
1S & 1.10236975[+01] &
   &                 &
   &                \\
2P & 1.05442022[+00] &
   &                 &
   &                \\
2P & 1.53765795[+00] &
   &                 &
   &                \\
2P & 2.24236213[+00] &
   &                 &
   &                \\
2P & 3.27003019[+00] &
   &                 &
   &                \\
3D & 1.64562346[+00] &
   &                 &
   &                \\
3D & 2.00041535[+00] &
   &                 &
   &                \\
3D & 2.43169940[+00] &
   &                 &
   &                \\
4F & 1.89664325[+00] &
   &                 &
   &                \\
4F & 2.36427250[+00] &
   &                 &
   &                \\
5G & 2.08330750[+00] &
   &                 &
   &                \\
\hline\hline
\end{tabular}
\end{table}

\begin{table}
\caption{Composition of STO basis sets for the beryllium atom. ETCC-5 basis sets contains all functions from the first
column. TC-ETCC-5 basis set contains additionally functions from the second column and ATC-ETCC-5 functions from the
third column. A-ETCC-5 contains functions from the first and third columns. The symbol $[k]$ denotes the powers of 10,
10$^k$.}
\begin{tabular}{cc|cc|cc}
\hline\hline
\multicolumn{2}{c|}{atomic valence} & 
\multicolumn{2}{c|}{tight core} & 
\multicolumn{2}{c}{diffuse} \\
\hline\\[-3.2ex]
1S & 5.56501242[$-$01] &
1S & 1.41661255[+01] &
1S & 4.33207693[$-$01] \\
1S & 8.08302238[$-$01] &
1S & 1.82673049[+01] &
2P & 6.71390454[$-$01] \\
1S & 1.17403603[+00] &
2P & 7.64626062[+00] &
3D & 7.58415000[$-$01] \\
1S & 1.70525397[+00] &
2P & 9.32939497[+00] &
4F & 8.96307000[$-$01] \\
1S & 2.47683292[+00] &
2P & 1.22141834[+01] &
5G & 1.02303000[+00] \\
1S & 3.59752943[+00] &
2P & 1.49624822[+01] &
6H & 1.15254000[+00] \\
1S & 5.22530925[+00] &
3D & 8.37670437[+01] &
   &                \\
1S & 7.58961318[+00] &
3D & 1.08645717[+01] &
   &                \\
1S & 1.10236975[+01] &
3D & 1.31895798[+01] &
   &                \\
2P & 1.02727422[+00] &
4F & 1.05215329[+01] &
   &                \\
2P & 1.37069132[+00] &
4F & 1.30025625[+01] &
   &                \\
2P & 1.82891254[+00] &
5G & 1.30737136[+01] &
   &                \\
2P & 2.44031682[+00] &
   &                 &
   &                \\
2P & 3.25611314[+00] &
   &                 &
   &                \\
3D & 1.61171171[+00] &
   &                 &
   &                \\
3D & 1.99156275[+00] &
   &                 &
   &                \\
3D & 2.46093774[+00] &
   &                 &
   &                \\
3D & 3.04093585[+00] &
   &                 &
   &                \\
4F & 1.88029525[+00] &
   &                 &
   &                \\
4F & 2.35223572[+00] &
   &                 &
   &                \\
4F & 2.94262984[+00] &
   &                 &
   &                \\
5G & 2.07200250[+00] &
   &                 &
   &                \\
5G & 2.84830932[+00] &
   &                 &
   &                \\
6H & 2.28502000[+00] &
   &                 &
   &                \\
\hline\hline
\end{tabular}
\end{table}

\begin{table}
\caption{Composition of STO basis sets for the beryllium atom. ETCC-6 basis sets contains all functions from the first
column. TC-ETCC-6 basis set contains additionally functions from the second column and ATC-ETCC-6 functions from the
third column. A-ETCC-6 contains functions from the first and third columns. The symbol $[k]$ denotes the powers of 10,
10$^k$.}
\begin{tabular}{cc|cc|cc}
\hline\hline
\multicolumn{2}{c|}{atomic valence} & 
\multicolumn{2}{c|}{tight core} & 
\multicolumn{2}{c}{diffuse} \\
\hline\\[-3.2ex]
1S & 5.56783742[$-$01] &
1S & 1.41615676[+01] &
1S & 4.33207693[$-$01] \\
1S & 8.08758496[$-$01] &
1S & 1.81731145[+01] &
2P & 6.71390454[$-$01] \\
1S & 1.17476545[+00] &
1S & 2.10111090[+01] &
3D & 7.58415000[$-$01] \\
1S & 1.70641035[+00] &
2P & 6.70939146[+00] &
4F & 8.96307000[$-$01] \\
1S & 2.47865332[+00] &
2P & 9.58760679[+00] &
5G & 1.02303000[+00] \\
1S & 3.60037799[+00] &
2P & 1.15480836[+01] &
6H & 1.15254000[+00] \\
1S & 5.22974374[+00] &
2P & 1.26805311[+01] &
7I & 1.28205000[+00] \\
1S & 7.59648561[+00] &
2P & 1.50027450[+01] &
   &                \\
1S & 1.10343062[+01] &
3D & 8.24507721[+00] &
   &                \\
1S & 1.60279264[+01] &
3D & 1.01353767[+01] &
   &                \\
2P & 1.02727422[+00] &
3D & 1.36819059[+01] &
   &                \\
2P & 1.37069132[+00] &
3D & 1.59970467[+01] &
   &                \\
2P & 1.82891254[+00] &
4F & 1.01588758[+01] &
   &                \\
2P & 2.44031682[+00] &
4F & 1.24513361[+01] &
   &                \\
2P & 3.25611314[+00] &
4F & 1.49991098[+01] &
   &                \\
2P & 4.34462965[+00] &
5G & 1.29310679[+01] &
   &                \\
3D & 1.61171171[+00] &
5G & 1.49929135[+01] &
   &                \\
3D & 1.99156275[+00] &
6H & 1.50083302[+01]  &
   &                \\
3D & 2.46093774[+00] &
   &                 &
   &                \\
3D & 3.04093585[+00] &
   &                 &
   &                \\
3D & 3.75762893[+00] &
   &                 &
   &                \\
4F & 1.88029525[+00] &
   &                 &
   &                \\
4F & 2.35223573[+00] &
   &                 &
   &                \\
4F & 2.94262984[+00] &
   &                 &
   &                \\
4F & 3.68120859[+00] &
   &                 &
   &                \\
5G & 2.07200250[+00] &
   &                 &
   &                \\
5G & 2.84830932[+00] &
   &                 &
   &                \\
5G & 3.91547113[+00] &
   &                 &
   &                \\
6H & 2.28502000[+00] &
   &                 &
   &                \\
6H & 3.42753000[+00] &
   &                 &
   &                \\
7I & 2.51517000[+00] &
   &                 &
   &                \\
\hline\hline
\end{tabular}
\end{table}

\begin{table}
\caption{Composition of the S-extended set of orbitals which are used for the relativistic calculations. See the main 
text for a more detailed description. The symbol $[k]$ denotes the powers of 10, 10$^k$.}
\begin{tabular}{c|c}
\hline\hline
orbital & exponent \\
\hline\\[-3.2ex]
1S & 5.30689678[$-$01] \\
1S & 7.58433830[$-$01] \\
1S & 1.08391382[$+$00] \\
1S & 1.54907273[$+$00] \\
1S & 2.21385343[$+$00] \\
1S & 3.16392312[$+$00] \\
1S & 4.52171284[$+$00] \\
1S & 6.46219464[$+$00] \\
1S & 9.23542936[$+$00] \\
1S & 1.31987909[$+$01] \\
1S & 1.88630191[$+$01] \\
1S & 2.69580367[$+$01] \\
1S & 3.85270110[$+$01] \\
1S & 5.50607817[$+$01] \\
1S & 7.86899790[$+$01] \\
\hline\hline
\end{tabular}
\end{table}

\setlength{\tabcolsep}{6pt}

\begin{table*}
\caption{Total energies of the beryllium dimer and atom calculated with frozen-core FCI method. $E_{HF}$ denotes the 
Hartree-Fock energy, $E_c$ denotes the correlation energy.  All values are given in the atomic units.}
\begin{ruledtabular}
\begin{tabular}{l|cccc}
\label{core}
          & \multicolumn{2}{c}{dimer} & \multicolumn{2}{c}{atom} \\
basis set & $E_{HF}$ & $E_c$ & $E_{HF}$ & $E_c$ \\
\hline\\[-2.1ex]
A-ETCC-2 & $-$29.1339418 & $-$0.1046873 & $-$14.5730210 & $-$0.0456697 \\
A-ETCC-3 & $-$29.1341621 & $-$0.1070574 & $-$14.5730222 & $-$0.0459687 \\
A-ETCC-4 & $-$29.1341745 & $-$0.1076392 & $-$14.5730231 & $-$0.0460547 \\
A-ETCC-5 & $-$29.1341751 & $-$0.1078505 & $-$14.5730231 & $-$0.0460901 \\
A-ETCC-6 & $-$29.1341754 & $-$0.1079423 & $-$14.5730231 & $-$0.0461091 \\
\end{tabular}
\end{ruledtabular}
\end{table*}

\begin{table*}
\caption{Total energies of the beryllium dimer calculated using various methods. These data are used for determination
of the inner-shell effects in the main text. All values are given in the atomic units. AE and FC are abbreviations for
all-electron and frozen-core, respectively.}
\begin{ruledtabular}
\begin{tabular}{lcccc}
\label{core}
method & ATC-ETCC-2 & ATC-ETCC-3 & ATC-ETCC-4 & ATC-ETCC-5 \\
\hline\\[-2.1ex]
AE CCSD & $-$29.314113 & $-$29.326129 & $-$29.329407 & $-$29.330683 \\[0.6ex]
FC CCSD & $-$29.234293 & $-$29.236054 & $-$29.236511 & $-$29.236681 \\
\hline\\[-2.1ex]
AE CCSDT & $-$29.319440 & $-$29.332239 & $-$29.335682 & $-$ \\[0.6ex]
FC CCSDT & $-$29.238668 & $-$29.240909 & $-$29.241460 & $-$ \\
\hline\\[-2.1ex]
AE CCSD(T) & $-$29.318577 & $-$29.331420 & $-$29.334925 & $-$29.3362862 \\[0.6ex]
FC CCSD(T) & $-$29.237825 & $-$29.240129 & $-$29.240743 & $-$29.2409768 \\
\hline\\[-2.1ex]
AE MP2 & $-$29.275033 & $-$29.290336 & $-$29.295338 & $-$29.297525 \\[0.6ex]
FC MP2 & $-$29.200976 & $-$29.204964 & $-$29.206567 & $-$29.207354 \\
\hline\\[-2.1ex]
AE MP4 & $-$29.311536 & $-$29.323856 & $-$29.327167 & $-$29.328456 \\[0.6ex]
FC MP4 & $-$29.230630 & $-$29.232484 & $-$29.232926 & $-$29.233094 \\
\end{tabular}
\end{ruledtabular}
\end{table*}

\begin{table*}
\caption{Total energies of the beryllium atom calculated using various methods. These data are used for determination
of the inner-shell effects in the main text. All values are given in the atomic units. AE and FC are abbreviations for
all-electron and frozen-core, respectively.}
\begin{ruledtabular}
\begin{tabular}{lcccc}
\label{core}
method & ATC-ETCC-2 & ATC-ETCC-3 & ATC-ETCC-4 & ATC-ETCC-5 \\
\hline\\[-2.1ex]
AE CCSD & $-$14.658664 & $-$14.663937 & $-$14.665410 & $-$14.665991 \\[0.6ex]
FC CCSD & $-$14.618818 & $-$14.619015 & $-$14.619088 &  \\
\hline\\[-2.1ex]
AE CCSDT & $-$14.659133 & $-$14.664551 & $-$14.666054 & $-$ \\[0.6ex]
FC CCSDT & $-$ & $-$ & $-$ & $-$ \\
\hline\\[-2.1ex]
AE CCSD(T) & $-$14.659105 & $-$14.664521 & $-$14.666028 & $-$14.666621 \\[0.6ex]
FC CCSD(T) & $-$14.618818 & $-$14.619015 & $-$14.619089 & $-$14.619120 \\
\hline\\[-2.1ex]
AE MP2 & $-$14.637838 & $-$14.644723 & $-$14.646999 & $-$14.647990 \\[0.6ex]
FC MP2 & $-$14.600887 & $-$14.602168 & $-$14.602758 & $-$14.603055 \\
\hline\\[-2.1ex]
AE MP4 & $-$14.655016 & $-$14.660288 & $-$14.661743 & $-$14.662320 \\[0.6ex]
FC MP4 & $-$14.614629 & $-$14.614731 & $-$14.614767 & $-$14.614789 \\
\end{tabular}
\end{ruledtabular}
\end{table*}